# Analysis of Supply Chain Network Using RFID Technique with Hybrid Algorithm

P Suresh and R Kesavan

**Abstract**— Radio Frequency IDentification (RFID) is a dedicated short range communication technology. The term RFID is used to describe various technologies that use radio waves to automatically identify people or objects. RFID is a method of remotely storing and retrieving data using RFID tag. Radio Frequency Identification (RFID) technology has been attracting considerable attention with the expectation of improved supply chain visibility for consumer goods, apparel, and pharmaceutical manufacturers, as well as retailers and government procurement agencies. RFID technology is used today in many applications, including security and access control, transportation and supply chain tracking. Supply Chain Management (SCM) is now at the centre stage of Manufacturing and service organizations. According to the strategies in markets, supply chains and logistics are naturally being modelled as distributed systems. The economic importance has motivated both private companies and academic researchers to pursue the use of operations research and management service tools to improve the efficiency of Transportation. Referring to such scenario, in this work RFID Technique adopted with hybrid algorithm to optimize supply chain distribution network.

——————————  ◆  ——————————

## 1 INTRODUCTION

Radio Frequency Identification (RFID) is a technology which allows contact less access to data on a transponder (also called tag or chip). Already in the late forties of the last century RFID was used to identify friendly aircraft. Ongoing miniaturization and advancements in technology have lead to smaller and cheaper tags, which have made widespread use of RFID possible in supply chains. The benefits of RFID in supply chains are well documented. Large retailers, e.g. Wal-Mart and the Metro Group as well as large consumer goods producers like Procter & Gamble and Unilever, are amongst the early adaptors of the technology. Defence Departments also expect significant efficiency gains and cost reduction for their Military Logistics Operations. Better stock keeping, reduced shrinkage, improved tracking, better information flows along the supply chain and a higher service level are some of the benefits attributed to the introduction of RFID. RFID is sometimes presented as a more sophisticated barcode or simply as the natural evolution from a paper-based to an electronic auto-ID technology. This analogy is dangerous, as it could result in inadequate risk management of RFID projects and systems. If not addressed, the specific nature of RFID, namely the wireless interface and the small computational footprint, might lead to security problems. The risks of RFID implementations are often solely seen as a consumer privacy problem, which can be dealt with at the point of sale by deactivating the tags. However, RFID specific security risks such as information leakage and data inconsistency arise along the entire supply chain. Ignoring the RFID specific risks in a supply chain environment can become quite costly. A preliminary consideration of the security risks is a prerequisite to achieve a successful RFID implementation. Rather surprisingly, the security implications of RFID projects for the supply chain are rarely addressed in a structured manner, but, if at all, on an ad hoc basis. First, we give a brief introduction to Supply Chain Management (SCM) and RFID, as well as to RFID specific supply chain setups. We explore the possible relations between begin and malicious RFID-readers as well as genuine and forged tags. The implementation difficulty and the benefits for the attacker build the base for the RFID risk classification. Finally we derive practical recommendations regarding risk management in RFID supported supply chains.

### A. RFID Systems

In a typical system tags are attached to objects. Each tag has a certain amount of internal memory (EEPROM) in which it stores information about the object, such as its unique ID (serial) number, or in some cases more details including manufacture date and product composition. When these tags pass through a field generated by a reader, they transmit this information back to the reader, thereby identifying the object. Until recently the focus of RFID technology was mainly on tags and readers which were being used in systems where relatively low volumes of data are involved. This is now changing as RFID in the supply chain is expected to generate huge volumes of data, which will have to be filtered and routed to the backend IT systems. To solve this problem

————————————

- *P Suresh is Assistant Professor, Department of Mechanical Engineering, Muthayammal Engineering College, Rasipuram, Namakkal District, Tamilnadu, India.*

- *R Kesavan is Assistant Professor, Department of Production Technology, Madras Institute of Technology, Anna University, Chromepet, Chennai, Tamilnadu, India.*

companies have developed special software packages called savants, which act as buffers between the RFID front end and the IT backend. Savants are the equivalent to middleware in the IT industry.

### B. RFID Tags

The heart of any RFID system is the tag, also known as the transponder. Two types of tags and their variants – active and passive – dominate the marketplace and are designed for specific applications. Active tags are battery powered with on-board battery power sources, contain up to hundreds of kilobytes of memory and can communicate data over longer ranges than passive tags. The tags are always on and always talking. Active tags, especially because of the built-in batteries, are far more expensive than passive tags. Active tags have been used for years and are typically used in long-range tracking applications and to track high-value inventory.

| ACTIVE TAGS | PASSIVE TAGS |
|---|---|
| Battery Powered | Beam Powered[1] |
| 100's of KB Memory | 64 — 256 bits Memory |
| Expensive | Inexpensive |
| Limited Battery Life | Long Life |
| Active Transmit | Backscatter[2] Transmit |
| Long Range | Short/Medium Range |
| Tag Talks First (TTF) | Reader Talks First (RTF) |

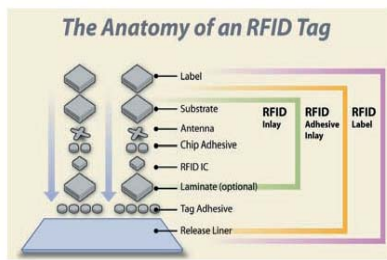

RFID-ANATOMY

### 2. SUPPLY CHAIN MANAGEMENT

The supply chain is a complex multi-stage process which involves everything from the procurement of raw materials used to develop products, and their delivery to customers via warehouses and distribution centres. Supply chains exist in service, manufacturing and retail organizations. Although, the complexity of the chain may vary greatly from industry to industry and firm to firm. Supply chain management (SCM) can be seen as the supervision of information and finances of these materials, as they move through the different processes, by coordinating and integrating the flows within and among the different companies involved. The efficiency of the supply chain has a direct impact on the profitability of a company. It is no surprise therefore to find that many large corporate companies have made it a key part of their strategy, and invested heavily in software systems (ERP, WMS.) and IT infrastructure designed to control inventory, track products and manage associated finance.

### A. Supporting Technologies for Supply Chain Management

For tracking & tracing resources along supply networks Auto-ID technology plays an important role. The Auto-ID technology developed by the Auto-ID Center consists of the following components combined in the EPC network
- **Radio Frequency Identification (RFID) tags and readers:** A tag is only a few square millimeters in size and is made up of a chip and an antenna. It can be scanned by special readers from a distance up to some meters dependent on the technology used.
- **Electronic Product Code (EPC):** EPC as the core of the EPC network is used for identification of physical objects in the real world. The Auto-ID Center released several different versions. The most common one is a 96-bit version with two different formats (General-Identifier and Serialized-General-Trade-Item-Number).
- **Object Naming Service (ONS):** ONS is a look-up service that delivers one or several IP addresses to the identified EPC.
- **Savant:** Savant acts as middleware that integrates the other components. The main functions are transmitting, filtering and bundling incoming data streams from readers to other services like ONS or existing enterprise applications.
- **Physical Markup Language (PML):** PML is a XML-based markup language to describe physical objects for standardized data exchange between the components of the EPC network.
- **EPC Information Service:** This service delivers product-related data of the observed objects.

### 2. APPLICATION OF RFID IN SCM

RFID provides a quick, flexible, and reliable electronic means to detect, identify, track, and manage a variety of items. The technology is well-suited for many operations in all types of industries provided that users develop new business processes to take advantage of RFID's special abilities. Merely substituting RFID for bar coding will not give users all the benefits that the technology could provide. Many potential users of RFID technology try to make comparisons between the relative cost of RFID and bar code, when comprehensive business process return on investment (ROI) analyses should be conducted. In recent years, interoperable products have emerged, helped by renewed standards efforts by EPC global and other standards bodies. In addition, major systems integrators have introduced RFID offerings and products that can make the adoption of RFID, especially in the enterprise, more straightforward as they build on existing and familiar systems. These efforts have enabled many companies to implement RFID pilots in their organizations and begin to calculate the ROI this technology can bring to their operations. The Proposed work chart diagram is given below.

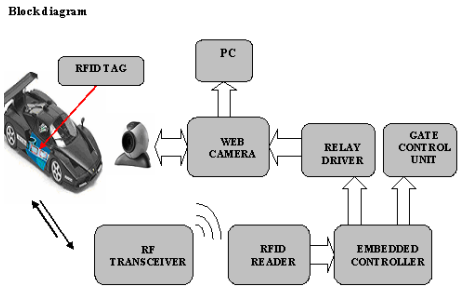

### A. Specifications

Embedded Controller:
Device Name            : P89C51RD2BN
Operating Voltage      : 4.5 – 5.5 V
ROM                    : 64KB
RAM                    : 1KB
Crystal Frequency      : 11.0592 MHz
Time taken for execution : 1.085µs
RFID Reader
Frequency              : 125 kHz
Operating Voltage      : 5V DC (± 5 %)
Read Range             : up to 3 cm
Serial Interface
Format                 : 9600Baud
No Parity, 8 Data bits, 1 Stop bit
Web Camera
I/O interface          : USB
Operating Voltage              : 3.3 V

## 3. OPTIMAL DESIGN OF SUPPLY CHAIN NETWORK

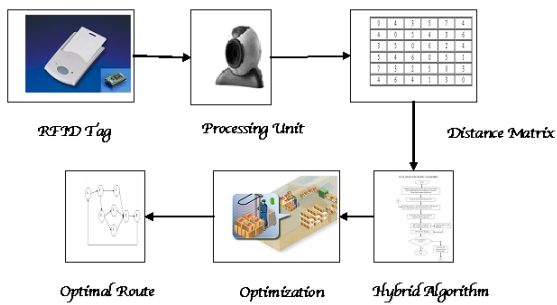

Fig. Proposed work of Design of optimal Supply Chain Network

The proposed work of optimization of Open loop Supply Chain Distribution network by using RFID techniques flow chart are also given below. Here, we have used the RFID tags to form the distance matrix of the all the customers distance. Then by using the random walker method used to solve the problem of ATSP. Then we are going the validate the results with the existing algorithm. The TSP has been studied with much interest within the last three to four decades. The majority of these works focus on the static and deterministic case of vehicle routing in which all information is known at the time of the planning of the routes. In most real –life applications though, stochastic and dynamic information occurs parallel to the routes being carried out. Real – life examples of stochastic and dynamic routing problems include the distribution of oil to private house holds, the pick-up of courier mail and packages and the dispatching of buses for the transportation of elderly and handicapped people. In these examples the customer profiles (i.e. the time to begin service, the geographic location, the actual demand etc.) may not be known at the time of the planning or even when service has begun for the advance request customers. Two distinct features make the planning of high quality routes in this environment much more difficult than it its deterministic counterpart; firstly, the constant change, secondly, the time horizon. A growing number of companies offer to service the customers within a few hours from the time the request is received. Naturally, such customer service oriented policies increase the dynamism of the system and therefore its complexity. During the past decade the number of published papers dealing with dynamic transportation models has been growing. The Hybrid Algorithm are given in below as flow chart.

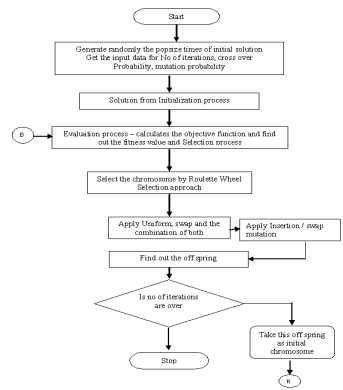

Fig. Flow chart of Hybrid Genetic Algorithm

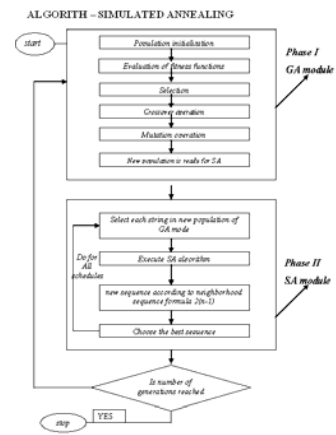

Fig. Flowchart of Hybrid GA-SA

A salesman has to visit n cities and return to his city of origin. Each city has to be visited exactly once, and the distance (or more generally the cost) of the journey between each pair of cities is known. The problem is to do the tour at minimum total cost. This forms one of a class of problems known as NP- complete problems, which are believed to require a computation time exp (kn), where k is a constant of the problem, and n is the number of cities for an exact solution. precisely, TSP can be interpret as follows: given a finite number of " cities" along with the cost of travel between each pair of them, find the cheapest way of visiting all the cities and returning to your starting point. A common special case of the TSP is the metric TSP, Where the distances between cities satisfy the triangle inequality. It means that there is little hope foe us to solve it in polynomial time unless P= NP. The proposed algorithm of Integrated Hybrid GA with SA is given below.

The TSP has been studied with much interest within the last three to four decades. The majority of these works focus on the static and deterministic case of vehicle routing in which all information is known at the time of the planning of the routes.  In most real –life applications though, stochastic and /or dynamic information occurs parallel to the routes being carried out. Real – life examples of stochastic and /or dynamic routing problems include the distribution of oil to private house holds, the pick-up of courier mail/packages and the dispatching of buses for the transportation of elderly and handicapped people. In these examples the customer profiles (i.e. the time to begin  service, the geographic location, the actual demand etc.) may not be known at the time of the planning or even when service has begun for the advance request customers. Two distinct features make the planning of high quality routes in this environment much more difficult than it its deterministic counterpart; firstly, the constant change, secondly, the time horizon. A growing number of companies offer to service the customers within a few hours from the time the request is received.  Naturally, such customer service oriented policies increase the dynamism of the system and therefore its complexity. During the past decade the number of published papers dealing with dynamic transportation models has been growing.

## 4. RESULT AND DISCUSSION

The performance of the algorithms GA, SA, Hybrid search algorithm combining GA and SA called Hybrid GA-SA algorithm, K-means Cluster based heuristic, Hybrid k-Means Cluster based heuristic-GA, Hybrid K-Means Cluster based heuristic- GA-SA for ATSP are evaluated by considering 14 pilot test problems. The problem data set is generated randomly for 14 numbers of cities and comparing with different optimization techniques validates the data set. The performance of the different algorithm is compared.

Table: Pilot Test Value

| SL. No | K-Means | GA | SA | GA-SA | K-GA-SA | K-GA | GA over SA | GA over GA-SA | GA-SA over SA | K-Means over SA | K-Means over GA-SA | K-Means GA Over SA | K-Means GA Over GA-SA | K-GA-SA over K-Means |
|---|---|---|---|---|---|---|---|---|---|---|---|---|---|---|
| 1 | 15620 | 20210 | 33400 | 22250 | 15320 | 14330 | 39.49102 | 9.168539 | 33.38323 | 22.71153 | 29.79775 | 8.258643 | 6.462141 | 1.920615 |
| 2 | 15660 | 21610 | 33360 | 21170 | 16690 | 14350 | 35.22182 | -2.07841 | 36.54077 | 27.53355 | 26.0274 | 8.365262 | 14.02037 | -6.57727 |
| 3 | 15700 | 21500 | 33120 | 21120 | 15760 | 15200 | 35.08454 | -1.79924 | 36.23188 | 26.97674 | 25.66288 | 3.184713 | 3.553299 | -0.38217 |
| 4 | 15720 | 21410 | 31940 | 20960 | 14400 | 15270 | 32.96807 | -2.14695 | 34.37696 | 26.57637 | 25 | 2.862595 | -6.04167 | 8.396947 |
| 5 | 15810 | 22010 | 34200 | 19830 | 15060 | 15780 | 35.64327 | -10.9934 | 42.01754 | 28.16901 | 20.27231 | 0.189753 | -4.78088 | 4.743833 |
| 6 | 15820 | 18560 | 31180 | 19840 | 15670 | 15650 | 40.47466 | 6.451613 | 36.36947 | 14.76293 | 20.2621 | 1.074589 | 0.127632 | 0.948167 |
| 7 | 15840 | 21000 | 33310 | 20960 | 19320 | 15600 | 36.95587 | -0.19084 | 37.07595 | 24.57143 | 24.42748 | 1.51512 | 19.25466 | -21.9697 |
| 8 | 15990 | 19590 | 32260 | 21080 | 15900 | 15430 | 39.27464 | 7.068311 | 34.65592 | 18.37672 | 24.14611 | 3.502189 | 2.955975 | 0.562852 |
| 9 | 16110 | 20040 | 31630 | 20660 | 14540 | 14970 | 36.64243 | 3.000968 | 34.68226 | 19.61078 | 22.01323 | 7.07635 | -2.95736 | 9.7455 |
| 10 | 16290 | 20340 | 32030 | 21060 | 16350 | 15380 | 36.49703 | 3.418603 | 34.24914 | 19.9115 | 22.64957 | 5.586249 | 5.932722 | -0.36832 |
| 11 | 16650 | 19280 | 33190 | 20330 | 13750 | 14720 | 41.91021 | 5.164781 | 38.74661 | 13.64108 | 18.10133 | 11.59159 | -7.05455 | 17.41742 |
| 12 | 16760 | 20400 | 32810 | 21000 | 14120 | 15430 | 37.82383 | 2.857143 | 35.99512 | 17.84314 | 20.19048 | 7.935561 | -9.27762 | 15.75179 |
| 13 | 16760 | 19720 | 33000 | 19170 | 14240 | 15930 | 40.24242 | -2.86907 | 41.90909 | 16.04014 | 12.57173 | 4.952267 | -11.868 | 15.0358 |
| 14 | 16820 | 21740 | 33720 | 19860 | 14560 | 13960 | 35.52788 | -9.46629 | 41.1032 | 22.63109 | 15.30715 | 17.00357 | 4.120879 | 13.43639 |

GA is compared with SA and Hybrid GA-SA, K-means Cluster based heuristic  is compared with GA, Hybrid GA- SA, Hybrid K-means Cluster based heuristic – GA and Hybrid K- Means Cluster based heuristic-GA- SA and Hybrid K-Means Cluster based heuristic –GA is compared with Hybrid K-Means Cluster based heuristic-GA-SA. In this work K-Means value arrived from the value generated by the RFID to the distance matrix. The comparative results are shown in graph below.

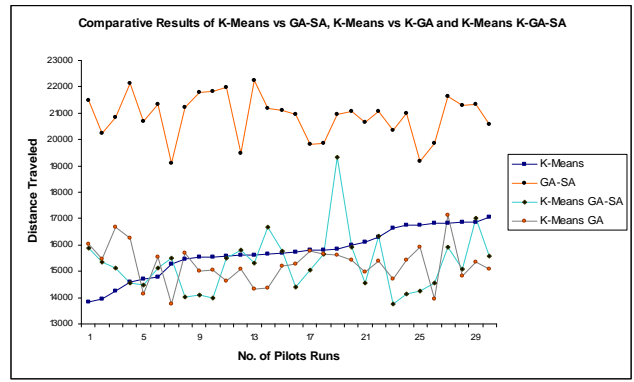

Fig. Comparative Results

## 5. CONCLUSION

RFID is a stable automatic identification technology that holds great promise for improving business processes; its use is becoming increasingly widespread. Indeed, some early adopter companies applying RFID at the carton and case level to mixed merchandise to automate creation of receiving and shipping manifests, have observed that RFID gives them unexpected opportunities to perform goods handling processes in efficient, entirely different ways. RFID should be considered for any application that could realize a clear benefit in terms of efficiency, reduced loss, or improved service. RFID offers strong performance and functionality, but at a price— considering tags relative to simple labels. The added cost of RFID, weighed against bar codes' outstanding value and the enormous installed, working infrastructure  ensures the two technologies will coexist, just as our nation's roads are still full of cars despite the growth of commercial air travel during the last 50 years. Because RFID tags can be reusable, don't require line of sight to read or write, enable unattended reading, and offer read/write data storage, they can improve efficiency in many operations by reducing labor and materials costs. Potential users must carefully evaluate the long-term impact for improved business operations relative to total cost of ownership and not automatically rule out use of the technology because of the initial investment required. As per the proposed algorithm, in this work RFID technique with Hybrid algorithm are adopted to optimize the Supply Chain Network ATSP. A simple but effective algorithms GA. SA, Hybrid GA-SA, K- Means Cluster based heuristic, Hybrid K- Means Cluster based heuristic-GA and Hybrid K- Means Cluster based heuristic – GA-SA for ATSP is proposed in this thesis. The problem of minimizing the distance travelled from centre single depot to all given n" cities are considered as the objective function for ATSP. GA has been modified by incorporating the uniform crossover operator and double operator crossover operator namely uniform crossover operator and swap crossover operator in the crossover process. The modified GA is compared with general GA and the modified is proved better obtaining nearer optimal solutions for ATSP.

**First Author**. **P. Suresh** received his B.E Degree in Mechanical Engineering from Madras University and Master Degree in Manufacturing Engineering from Madras Institute of Technology, Anna University, Chennai. And also he received his MBA Degree from Bharathiar University, Coimbatore. He is doing Ph.D in Supply Chain Management at Anna University, Chennai. He is at present working as an Assistant Professor in the Department of Mechanical Engineering, Muthayammal Engineering College, Rasipuram, Namakkal District, Tamilnadu. His field of interest is Supply chain Management, Ergonomics, Image Processing, Bio-Mechanics, Industrial Engineering, Production and Operations Management and Manufacturing Engineering.

**Second Author.** **Dr.R.Kesavan** received his B.E degree in Mechanical Engineering and M.E Degree in Production Engineering from College of Engineering, Guindy. And also he received his MBA Degree in Madras University. He also received his Ph.D in Ergonomics from Anna University, Guindy, Chennai. He is at present working as a Assistant Professor, Department of Production Technology, MIT Campus, Anna University, Chennai. His field of interest is Ergonomics, TQM, Production and operations Management, Supply Chain Management, Logistics.